# Two dimensional characterization of space-momentum entangled photon pairs

**Martin Ostermeyer, Dietmar Korn, Dirk Puhlmann**
*University of Potsdam, Institute of Physics and Astronomy, Karl-Liebknecht-Str. 24/25, 14476 Potsdam, Germany*
[*]*Corresponding author: oster@uni-potsdam.de*

**Abstract**: Space momentum entangled photon pairs are generated from type II parametric down conversion in a beta barium borate crystal. The correlations in the positions of photons in the near field and far field planes with regard to the generating crystal are observed in both transverse dimensions using scanning fiber probes. The space-momentum correlation is characterized using a covariance description for a bivariate normal distribution and tested for non-separability with Mancini's criterion. The role of higher order spatial modes to observe spatial entanglement between the two photons is discussed.

OCIS codes:.

## 1. Introduction

Conceptual applications of space-momentum entanglement go as far back as the Gedanken experiment of Einstein, Podolski, and Rosen (EPR) [1]. Besides the discussion of adequate interpretation of quantum mechanics and its completeness, a wide range of applications of entanglement in the area of quantum information processing evolved. Spatial entanglement opened further possibilities for advanced methods in quantum imaging [2, 3]. Some of these concepts, e.g. quantum lithography [4] or entangled N-photon microscopy [5, 6], rely on N-photon absorption. For the here discussed photon pairs (N=2) from an entangled photon source, the two-photon absorption rate depends not only on the probability that both photons arrive within a period smaller than the virtual state lifetime, but also on the probability that the two photons arrive within an area smaller than the absorption cross section [7, 5]. Thus, detailed knowledge of the spatial correlation of potential sources is vital for some of the quantum imaging methods. Furthermore, the spatial structure of space-momentum entanglement from parametric down conversion provides new possibilities for exploring correlations in the non-classical domain beyond the two dimensional Hilbert space of polarization entanglement.

There are investigations of spatial entanglement related to our investigation in the literature. Recently, Fedorov et. al. [8] discussed the anisotropy of the spatial biphoton correlation from type I parametric down conversion in the far field of the generating crystal. They found a strong narrowing of the coincidence profile compared to the single photon mode width by a factor of 80 in the plane containing optical and laser axis. Entanglement between photons from parametric down conversion was experimentally considered and used in imaged near and far-field planes in experiments in the group of Y.Shi [9] with reference to the Gedankenexperiment of Popper [10] and R. Boyd that approximate EPR-states to relate to the EPR-paradox (see e.g. [11] and references therein). None of these investigations aimed for a complete spatial mapping of the correlations between the photons from SPDC in their position and momentum in the two transverse dimensions. In the experiments reported in [11], the correlation of biphotons from parametric down conversion was investigated in two conjugated planes. Measurements were performed in one spatial dimension with one detector fixed in one central position and the other one scanned along a line. In this paper we proceed one step further and evaluate the correlation between the photons at arbitrary positions $x_1$ and $x_2$ and arbitrary momenta $p_{x,1}$ and $p_{x,2}$ in the transverse dimensions. Due to detector scans on both modes of the bipartite setting in our investigation, the two-photon position separation $\Delta(x_1-x_2)$ and the two-photon momentum sum $\Delta(p_{x,1}+p_{x,2})$, as precisely used in the criterion by Mancini et. al. in the context of opto-mechanical coupling, can be derived. Moreover, the complete mapping we report on allows for further characterization of the space momentum entanglement using covariance matrices [12], enabling easy comparison with other entangled continuous variable systems.

## 2. Characterize entanglement in continuous variables

The variables of interest that are addressed in the following are always the transverse properties of the photon's position and momentum. Entanglement in position and momentum as entanglement in any pair of canonical coordinates is entanglement in continuous variables. There are different proposals to characterize the degree of this kind of entanglement with appropriate measures, which can also be detected using appropriate criteria based on second moments. This degree of entanglement can for example be quantified in terms of the negativity, which is in fact an entanglement monotone and hence a proper measure of entanglement [13, 14, 15]. For pure states, the Schmidt number – the rank of the reduced state - also reasonably quantifies the entanglement content. Interestingly, for multi-mode systems of light, the number of relevant and significantly populated Schmidt modes gives rise to a clear picture of how many modes effectively contribute to the bipartite entanglement (see [16, 17]). When one aims at merely detecting but not quantifying entanglement, a characterization is experimentally readily accessible by means of the ratios of the standard deviations of the single and coincidence rate distributions in either near or far field [18, 19, 20].

For our characterization, and in order to detect entanglement in the system under consideration, we make use of the criterion of ref. [22], in the variant of a product criterion of ref. [21]. These criteria, and in fact all criteria which are linear or quadratic in second moments, can easily be expressed and completely characterized in terms of covariance matrices [23, 12]. Applied to space and momentum, the product criterion states that for any bipartite quantum state with the non-commuting observables $x_1$ and $p_1$ for the first particle and $x_2$ and $p_2$ for the second particle, the limit for the non-separability of the state is reached, when their product of variances satisfies

$$\Delta^2(x_1 - x_2)\,\Delta^2(p_{x,1} + p_{x,2}) < \hbar^2. \qquad (1)$$

If the measured data is detailed enough, covariance matrices can be used to characterize the correlation and to check for separability of the state. The full covariance matrix for our two-mode-setting in position and momentum reads:

$$\Sigma_{x_1 x_2 p_{x,1} p_{x,2}} = \begin{pmatrix} \text{Var}(x_1) & \text{Cov}(x_1, x_2) & \text{Cov}(x_1, p_{x,1}) & \text{Cov}(x_1, p_{x,2}) \\ \text{Cov}(x_2, x_1) & \text{Var}(x_2) & \text{Cov}(x_2, p_{x,1}) & \text{Cov}(x_2, p_{x,2}) \\ \text{Cov}(p_{x,1}, x_1) & \text{Cov}(p_{x,1}, x_2) & \text{Var}(p_{x,1}) & \text{Cov}(p_{x,1}, p_{x,2}) \\ \text{Cov}(p_{x,2}, x_1) & \text{Cov}(p_{x,2}, x_2) & \text{Cov}(p_{x,2}, p_{x,1}) & \text{Var}(p_{x,2}) \end{pmatrix} \qquad (2)$$

$$= \begin{pmatrix} \Sigma_{x_1 x_2} & \text{not used} \\ \text{not used} & \Sigma_{p_{x,1} p_{x,2}} \end{pmatrix}$$

For our simultaneously measured observables position $x_1$, $x_2$ and momentum $p_{x,1}$, $p_{x,2}$ respectively, the covariance sub-matrices with subtracted first moments of the two modes 1 and 2 are of interest:

$$\Sigma_{x_1 x_2} = \begin{pmatrix} \text{Var}(x_1) & \text{Cov}(x_1, x_2) \\ \text{Cov}(x_2, x_1) & \text{Var}(x_2) \end{pmatrix} = \underbrace{\begin{pmatrix} \sigma_{x_1}^2 & \rho_x \sigma_{x_1} \sigma_{x_2} \\ \rho_x \sigma_{x_2} \sigma_{x_1} & \sigma_{x_2}^2 \end{pmatrix}}_{\text{for Gaussian normal distribution}} \quad (3a)$$

$$\Sigma_{p_{x,1} p_{x,2}} = \begin{pmatrix} \text{Var}(p_{x,1}) & \text{Cov}(p_{x,1}, p_{x,2}) \\ \text{Cov}(p_{x,2}, p_{x,1}) & \text{Var}(p_{x,2}) \end{pmatrix} = \underbrace{\begin{pmatrix} \sigma_{p_{x,1}}^2 & \rho_{p_x} \sigma_{p_{x,1}} \sigma_{p_{x,2}} \\ \rho_{p_x} \sigma_{p_{x,1}} \sigma_{p_{x,2}} & \sigma_{p_{x,2}}^2 \end{pmatrix}}_{\text{for Gaussian normal distribution}} \quad (3b)$$

The off-diagonal matrices (labeled as not used) are not considered any further. The elements of the sub-matrices in eq. (3a) and (3b) are sufficient to express the criterion of Mancini and hence to decide if the considered two-photon state is entangled. Thus, we did not measure any data to calculate the elements of the off-diagonal matrix elements

The covariance formalism shall be applied to our measured data. The distribution of acquired data on position and momentum (see section 3) has been fitted with Gaussian functions since the field of view of the observed two dimension mode cross sections is limited. Consequently, the covariance matrix above is expressed for the resulting bivariate normal distribution by the standard deviations of the distributions of the two modes $\sigma_{x_1}$, $\sigma_{x_2}$ and $\sigma_{p_{x,1}}$, $\sigma_{p_{x,2}}$ respectively. $\rho_x$ and $\rho_{p_x}$ quantify the correlation strength in the statistical sense. They range from -1 to 1. The uncorrelated case is expressed by $\rho = 0$ and the correlated and anti-correlated cases by 1 and -1 respectively. Using this covariance matrix the probability to detect one photon at $x_1$ and the other at $x_2$ can be written as:

$$p_{\text{Cov}}(x_1, x_2) = \frac{1}{2\pi \sqrt{|\Sigma_{x_1 x_2}|}} \exp\left(-\frac{1}{2} (x_1 \; x_2) \Sigma_{x_1 x_2}^{-1} \begin{pmatrix} x_1 \\ x_2 \end{pmatrix}\right) \quad (4a)$$

$$\Sigma_{x_1 x_2}^{-1} = \frac{1}{1-\rho_x^2} \begin{pmatrix} \frac{1}{\sigma_{x_1}^2} & \frac{-\rho_x}{\sigma_{x_1} \sigma_{x_2}} \\ \frac{-\rho_x}{\sigma_{x_2} \sigma_{x_1}} & \frac{1}{\sigma_{x_2}^2} \end{pmatrix} \quad (4b)$$

For the correlation of photons from parametric down conversion, new coordinates s and t are suitable:

$$s = \frac{x_1 + x_2}{2} \qquad t = \frac{x_2 - x_1}{2} \quad (5a)$$

To meet Mancini's criterion, slightly different coordinates u and v with a different metric are used:

$$u = x_1 + x_2 \qquad v = x_1 - x_2 \quad (5b)$$

These coordinates denote the diagonal axis of the $x_1$-$x_2$-coordinate system. s and u express the direction along which a two-photon detector (consisting of two single photon detectors at identical position $x_1 = x_2$) would be scanned to measure the variance of the correlation. t and v express the direction, along which two detectors with $x_2 = -x_1$ would be scanned, to measure the variance of the anti-correlation. In case of s and t the x-coordinates are divided by two to conserve the metric of the two photon detector scan.

For the special, but here expected case of equal standard deviations of the two modes $\sigma_{x_1} = \sigma_{x_2} = \sigma_{in}$, the following covariance matrices are obtained:

$$\Sigma_{x_1 x_2} = \sigma_{in}^2 \begin{pmatrix} 1 & \rho_x \\ \rho_x & 1 \end{pmatrix} \qquad \Sigma_{x_1 x_2}^{-1} = \frac{1}{(1-\rho_x^2) \sigma_{in}^2} \begin{pmatrix} 1 & -\rho_x \\ -\rho_x & 1 \end{pmatrix} \quad (6a)$$

$$\Sigma_{st} = \frac{\sigma_{in}^2}{2}\begin{pmatrix} 1+\rho_x & 0 \\ 0 & 1-\rho_x \end{pmatrix} \qquad \Sigma_{st}^{-1} = \frac{2}{\sigma_{in}^2}\begin{pmatrix} \frac{1}{1+\rho_x} & 0 \\ 0 & \frac{1}{1-\rho_x} \end{pmatrix}. \tag{6b}$$

So that the probability distribution transforms to:

$$p_{\text{Cov}}(x_1, x_2) \propto \exp\left(-\frac{1}{2\sigma_{in}^2(1-\rho_x^2)}(x_1^2 + x_2^2 - 2\rho_x x_1 x_2)\right) \tag{7a}$$

$$p_{\text{Cov}}(s, t) \propto \exp\left(-\frac{1}{2\sigma_{in}^2}\left(\frac{2s^2}{1+\rho_x} + \frac{2t^2}{1-\rho_x}\right)\right) \tag{7b}$$

$$p_{\text{Cov}}(u, v) \propto \exp\left(-\frac{1}{2\sigma_{in}^2}\left(\frac{u^2}{2(1+\rho_x)} + \frac{v^2}{2(1-\rho_x)}\right)\right) \tag{7c}$$

These probability distributions can be illustrated in correlation diagrams. The probability to detect two photons at $x_1$ and $x_2$ is proportional to the coincidence rate at the positions of the two photons $x_1$ and $x_2$. This rate is depicted as a pixel contour plot, according to our resolution in the measurements described below, with the position of mode one and two as variables for the abscissa and ordinate. For uncorrelated photons with $\rho_x = 0$ in a mode with a Gaussian transverse intensity distribution, as can be observed e.g. in a coherent beam, such a correlation diagram is shown in Fig. 1. The distribution is centered around the first moment of the Gauss-distribution and has isotropic spherical shape. In contrast to such an isotropic shape, correlated photons will be found to exhibit an elliptical probability distribution in the correlation diagram. The perfect correlation of the photons generated in the same location with $\rho_x = 1$ would result a line along $x_2 = x_1$ with infinitely small width. For photons from parametric down conversion with correlated photons in the near field and anti-correlated photons in the far field one would expect correlation diagrams as shown in Fig. 2.

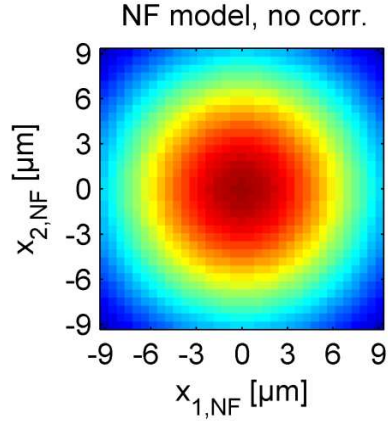

*Fig. 1: Calculated correlation diagram of uncorrelated photons, as e.g. in ordinary coherent light detected in a spatially resolved Hanbury-Brown and Twiss setup (see Fig. 4, red denotes high intensity).*

The standard deviation of the correlation, if only one detector e.g. at $x_1$ is moved, would be evaluated along a horizontal line in the correlation diagram and results in case of $x_2 = 0$ to

$$\sigma_{x_1|(x_2=0)} = \sigma_{in}\sqrt{(1-\rho_x^2)} \tag{8}$$

This yields the identical standard deviation as for one of the modes in case of no correlation ($\rho_x = 0$) and a standard deviation of zero in case of perfect correlation ($\rho_x = 1$).

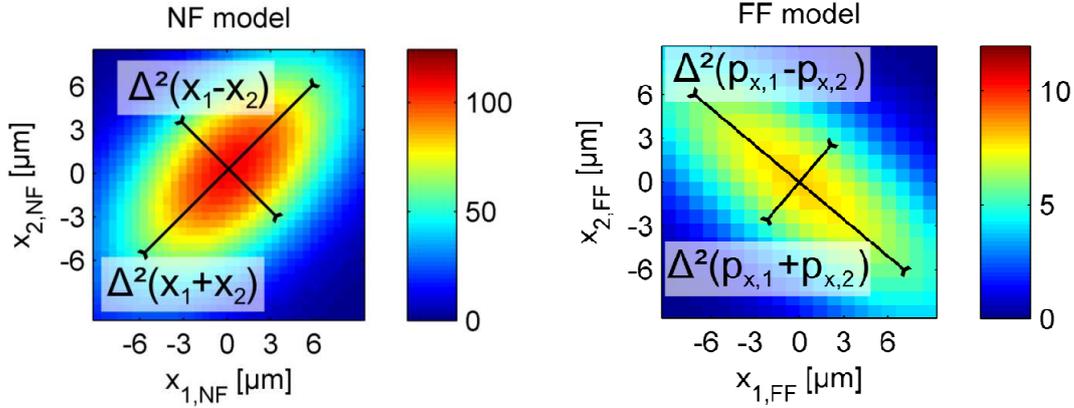

*Fig. 2: Calculated correlation diagram of coincident, correlated photons as expected in the near field (left) and far field (right) of the source detected in a spatially resolved Hanbury-Brown and Twiss setup (see Fig. 4). Coincidence rate in arbitrary units.*

The standard deviations of the position along s and t direction are directly obtained following eq. 5a (analog expressions for the momenta in the s-t-coordinates):

$$\sigma_{s,x}(\rho_x) = \sigma_{in}\sqrt{\frac{1+\rho_x}{2}} = \Delta\left(\frac{x_1+x_2}{2}\right) \qquad \sigma_{t,x}(\rho_x) = \sigma_{in}\sqrt{\frac{1-\rho_x}{2}} = \Delta\left(\frac{x_2-x_1}{2}\right) \qquad (9)$$

This means along t-direction a standard deviation reduced by a factor of $\sqrt{2}$ in case of no correlation and a standard deviation of zero in case of perfect anti-correlation ($\rho_x = -1$) or identical standard deviation of the single photon case in case of perfect correlation ($\rho_x = 1$). Along s-direction, the standard deviation $\sigma_{s,x}$ equals $\sigma_{t,x}$ in the uncorrelated case, but for growing correlation strength it approaches the original standard deviation $\sigma_{in,x}$ from the single photon case, whereas in the anti-correlation case it approaches zero.

Expressing Mancini's criterion in the u-v-coordinate system using the covariance parameters $\rho_x$ and $\rho_p$ the expression reads:

$$2\sigma_{in,x}^2(1-\rho_x)2\sigma_{in,p_x}^2(1+\rho_{p_x}) < \hbar^2 \qquad (10)$$

The product of the standard deviations in x and p gives a measure for the number of higher order spatial modes that are involved in the state of light that is investigated. The mode volume of a phase cell is $\sigma_{x,00} \cdot \sigma_{p,00} = \hbar/2$. In particular this is the case for the $TEM_{00}$. Within the laser community a coherent beam of light that obeys this condition is called diffraction limited. Partially coherent beams can be characterized by a so called $M^2$ number (see e.g. [27]). In general their phase volume is $M^2$ times that of the phase cell: $\sigma_x \cdot \sigma_p = M^2 \hbar/2$. Using this $M^2$-number, an effective indicator for the spatial multimode character of bipartite entanglement is given.

Using these relations the covariance criterion for non-separability can be expressed as:

$$(1-\rho_x)(1+\rho_{p_x})(M^2)^2 < 1 \qquad (11)$$

On the other hand, restricting the generation and propagation of higher order spatial modes leads to a lower bound of the observable strength of correlation. This limitation can be considered in two different ways.

The first consideration directly uses the uncertainty relations for the non-commuting observable pairs $\Delta(x_1 + x_2)\cdot(p_{x,1} + p_{x,2}) > \hbar$ (a) and $\Delta(x_1 - x_2)\cdot \Delta(p_{x,1} - p_{x,2}) > \hbar$ (b).

If $\Delta(p_{x,1} + p_{x,2})$ ( should be small then following (a), $\Delta(x_1 + x_2)$ has to be allowed to become large. Assuming a perfect spatial correlation with $x_1 = x_2$, $\Delta(p_{x,1} + p_{x,2})$ (is limited by the emission area in the nonlinear crystal with radius $w_{1,2}$ (second moment definition of beam radii see e.g. [27], $j = 1,2$) that is equal to the pump spot size $w_p$ to

$$\Delta(p_{x,1} + p_{x,2}) \geq \frac{\hbar}{w_p} = \frac{\hbar}{w_j} \quad (12a)$$

If $\Delta(x_1 - x_2)$ should be minimized, $\Delta(p_{x,1} - p_{x,2})$ on the other hand has to be allowed to become large. Using the phase matching condition $p_{x,1} + p_{x,2} = 0$ and following (b) the limitation becomes

$$\Delta(x_1 - x_2) \geq \frac{1}{k_{x,j}\,\Theta_{x,j}}. \quad (12b)$$

$k_{x,j}$ is the wave number of the signal and idler photons respectively in the degenerate case. $\Theta_{x,j}$ denotes the divergence of the signal and idler photons.

Second, the standard deviations $\Delta(x_1 - x_2)$ and $\Delta(p_{x,1} + p_{x,2})$ can be considered directly. Because of momentum conservation $\Delta(p_{x,1} + p_{x,2}) = \Delta(p_{x,\text{pump}})$, the quantity $\Delta(p_{x,\text{pump}})$ in turn is connected to the far field divergence $\Theta_{\text{pump}}$ of pump beam $\Delta(p_{x,\text{pump}}) = \frac{\hbar k_{\text{pump}}\Theta_{\text{pump}}}{2}$, which leads to

$$\Delta(p_{x,1} + p_{x,2}) \geq \frac{\hbar k_{\text{pump}}\Theta_{\text{pump}}}{2}. \quad (12c)$$

The two photons in the near field can be distinguished with regard to their position if they are not within one phase cell which means

$$\Delta(x_1 - x_2) \geq \frac{1}{k_{x,j}\,\Theta_{x,j}}. \quad (12d)$$

Condition 12a and 12c imply that pumping with a large diameter low divergence beam of low spatial mode order, to minimize both quantities simultaneously, is beneficial on one hand. On the other hand, 12b and 12d in combination with 12a imply, since $w_p = w_j$, that for the generated signal and idler photons a high order spatial multimode case with both large $w_{s,i}$ and large $\Theta_{s,i}$ should be allowed for to minimize $\Delta(x_1 - x_2)$ and $\Delta(p_{x,1} + p_{x,2})$ simultaneously. Condition 12a and 12d can be combined to give a lower bound for the non-separability criterion of eq. 11:

$$\frac{1}{(2M_j^2)^2} \leq \left(1 - \rho_x(M_j^2)\right)\left(1 + \rho_{p_x}(M_j^2)\right)(M_j^2)^2 < 1 \quad (13)$$

Since this is a first order approximation with Gaussian and hard apertures to restrict the propagating modes, there will be an uncertainty of the factor $1/(2M^2)^2$ in the range of $(1/2^2)^2$ and the absolute value of this lower bound can only be given within some margin. Still, this criterion illustrates in principle that the observation of higher order spatial modes is beneficial to push the lower bound concerning the correlation limitation further down. This result is quite similar to the result of van Exter et. al. [28]. They compare the meaning of an acceptance mode number (Etendue) of the optical setup, used to observe the two-photon state, and the two-photon Schmidt number. But on the other hand, since $\rho = (\sigma_s^2 - \sigma_t^2)/(\sigma_s^2 + \sigma_t^2)$, the absolute value of $\rho$ mirrors the level of two mode squeezing of the correlation ellipse. According to eq. 13, this relative squeezing of the ellipse needs to be stronger in presence of higher order modes to result in the equal distance of the product in eq. 13 from the non-separability bound on the right hand side of the equation. To get deeper insight in the spatial mode dynamics of the space momentum entanglement, ρ has to be formulated for the specific spatial modes emitted by the parametric down conversion source which goes beyond the scope of this investigation.

## 3. Setup to characterize a biphoton beam

We generate correlated photons from noncollinear type II parametric down conversion [24] in a β-barium borate (BBO) crystal. The schematic of the setup is shown in Fig. 3. The BBO crystal is pumped by a frequency doubled modelocked ps Tisa-laser with a repetition rate of 76 MHz and 20mW average power at a wavelength of 390 nm. The pump spot diameter is 160 µm. The run time differences due to the birefringence of the BBO are compensated to first order by a half wave plate and compensation BBO-crystals of half thickness [24]. The two photons hit a 50 % beam splitter, where by Hong-Ou-Mandel (HOM) interference [29] they reunite to two biphoton beams at the upper and lower port of the beam splitter. The visibility of the HOM-dip is 95 %. The pairs are detected with a coincidence rate of 3000 counts/s.

We have chosen a non-collinear type II down conversion process to have the possibility to manipulate the photons independently before there are interfered to produce the biphoton beam. (The full possibility of this independent manipulation is not exploited in the work reported here.) The interference at the beam splitter is adjusted for maximum mode overlap by an optimization of the visibility of the HOM-dip. The photons are detected by single mode fiber coupled single photon detectors (SPCM-AQRH 15 by PerkinElmer).

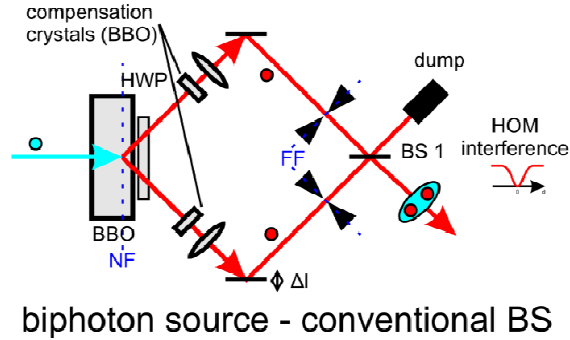

*Fig. 3: Schematic setup to generate a biphoton beam by Hong-Ou-Mandel interference of photons produced from type II non-collinear parametric down conversion in -Barium-Borate (BBO).*

We use a Hanbury-Brown and Twiss like setup to measure the spatial correlations of the two photons. The reference plane is always the nonlinear crystal front facet. To investigate the correlations of the photons, the near field plane of the crystal is imaged with a telescope of lenses with 400 mm and 60 mm (achromat) focal length respectively (see Fig. 4). The far field of the crystal is realized in the focal plane of a f = 400 mm lens and is demagnified with a telescope consisting of a two lenses with focal lengths of 1000 mm and 11 mm (asphere) resulting in a demagnification factor of K = 226.4 m. The transverse momentum is derived from the measured coordinates of the fiber probe position $x_{1,FF}$ and $x_{2,FF}$: $p_{x,1} = \hbar k \cdot x_{1,FF}/K$.

The area that is scanned is 20 µm x 20 µm with a resolution of 35 by 35 steps in the imaged near field plane and 20 by 20 steps in the imaged far field plane. The single mode fibers have a mode field diameter of 5.3 µm.

During the typical measurement procedure, fiber probe one (passive probe) was set to a fixed position and the other fiber (active probe) scanned along a line. E.g. for a characterization in the x-dimension both fibers are held in central y-positions (y=0).

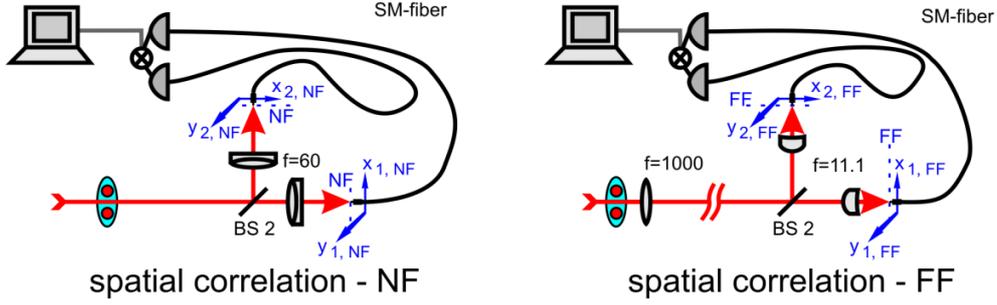

*Fig. 4: Schematic setup to measure the distribution of the coincidence rates in the biphoton beam in the imaged near field and the imaged far field of the BBO crystal.*

The single and coincidence rates are recorded for each step along the scan. Then the passive fiber was moved to the next x-position, fixed there, before the active fiber probe was scanned along the same line again. We performed scans in the imaged near field and in the imaged far field planes. Since the single photon rates are low, the possibility to detect photons that belong to different pairs is negligible. During these measurements we obtained coincidence count rates of about 100 counts/s in the near field plane and 10 counts/s in the far field plane.

The result of an additional measurement, where a complete 2-dimensional x-y-scan of the active fiber probe across the mode was carried out for thirteen different positions of the passive fiber, is depicted in Fig. 5 (evaluation of seven positions is shown). In contrast to Fig. 1 and Fig. 2 the two dimensional contour plots here are real spatial two dimensional cross sections along the transverse vertical and horizontal direction. Although the position of the single photon mode cross section remains the same for each passive fiber probe position, as expected due to the common origin of the two photons of one biphoton, the center of the coincidence distribution moves from left to right in case of the horizontal scan and from top to bottom in case of the vertical scan in parallel with the change in position of the passive fiber stage. The quantitative evaluation of this observation is carried out in the next section.

## 4. Evaluation and illustration of experimental data

The measurement results are evaluated and illustrated in correlation diagrams as discussed in section 2. The coincidence rate of the measurements is depicted as contour plot with the position of the fiber probe one and two as parameters for the abscissa and ordinate as discussed in the context of Fig. 1 and Fig. 2. The results of the measurements in the imaged near and far field plane of the x-dimension are shown in Fig. 6. The spot size of the single photon intensity in the imaged far field plane was bigger compared to the imaged near field plane. To get the measurement scans done under stable conditions, we reduced the number of scanning steps for the far field measurements. The measured data points were fitted with a two-dimensional Gaussian function, in which in contrast to eqs. 7abc the direction of the main axis was not fixed but given by an angle α:

$$R_{c,\text{Bell}}(x_1, x_2; x_{1_0}, x_{2_0}, \alpha, r, \sigma_m, \sigma_n) = r \exp\left(\frac{-x_m^2}{2\sigma_m^2} + \frac{-y_n^2}{2\sigma_n^2}\right) \quad (14)$$
$$x_m = (x_1 - x_{1_0}) \cos\alpha - (x_2 - x_{2_0}) \sin\alpha$$
$$y_n = (x_1 - x_{1_0}) \sin\alpha + (x_2 - x_{2_0}) \cos\alpha$$

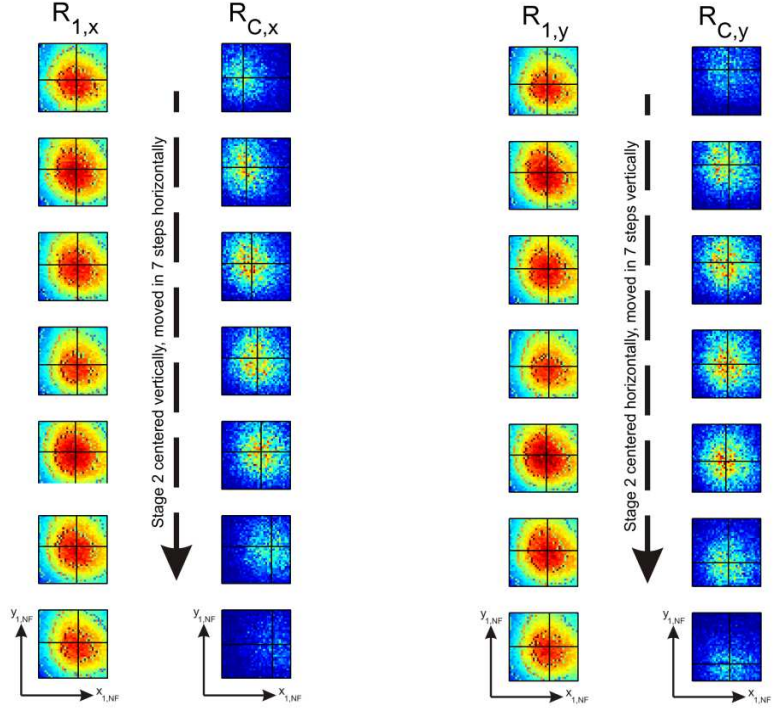

*Fig. 5: Measured 2-dimensional spatial distributions for a horizontal x-scan (first two colums) and a vertical y-scan (last two colums).Shown is in each case the single photon count rate of the active fiber probe (first and third column) and the coincidence rate (second and fourth column). For the horizontal scan (vertical scan) the passive fiber stage was moved to seven different positions in x-direction (y-direction) and centered in y-direction (x-direction) and fixed for the scan of the active probe. The coincidence distributions are shown in dependence of the spatial scan coordinates of the active probe.*

$$R_{c,\text{Cov}}(x_1, x_2; x_{1_0}, x_{2_0}, \rho_x, r, \sigma_{x_1}, \sigma_{x_2}) = r\exp\left(-\frac{1}{2}v_{x_1x_2}^T \Sigma_{x_2x_2}^{-1} v_{x_1x_2}\right) \quad (15)$$

$$v_{x_1x_2} = \begin{pmatrix} x_1 - x_{1_0} \\ x_2 - x_{2_0} \end{pmatrix} \quad \Sigma_{x_1x_2} = \begin{pmatrix} \sigma_{x_1}^2 & \rho_x \sigma_{x_1} \sigma_{x_2} \\ \rho_x \sigma_{x_2} \sigma_{x_1} & \sigma_{x_2}^2 \end{pmatrix}$$

The resulting fit function is plotted in the same diagram by contour lines.
The fits result in an angle α = 44° in the near field and α = -51° in the far field which is close to +45°/-45°. α = 45° indicates a major axis along $x_2 = x_1$ which means a common position of origin of the two photons whereas α = -45° indicates the direction of $p_{x,1} = -p_{x,2}$, the momentum anti-correlation. The standard deviation $\sigma_u$ and $\sigma_v$ needed for Mancini's criterion in the case of α = +45°/-45° result from the fit data as $\sigma_u = \sqrt{2}\sigma_m$, and $\sigma_v = \sqrt{2}\sigma_n$.
The two dimensional fit of the measured data results in a variance product in x-direction of

$$\Delta^2(x_1 - x_2) \cdot \Delta^2(p_{x,1} - p_{x,2}) = 0.16 \cdot \hbar^2$$

Using the covariance matrix the evaluation yields $\sigma_{x,\text{in}} = 39.7$ μm, $\sigma_{px,\text{in}} = 15300$ $\hbar$/m and correlation coefficients of $\rho_x = 0.53$ and $\rho_{p_x} = -0.77$ leading to a product on the left hand side of eq. 10 of $0.16\ \hbar^2$.
The measured correlation diagrams in y-dimension look similar to the ones in x-dimension. The fluctuations are somewhat bigger, and the resulting correlation strength is somewhat smaller. Since BBO is an uniaxial crystal, asymmetries in the strength of the correlation between the different dimensions are to be expected as discussed in [8]. In our case, the alignment for the HOM interference might be less perfect in y than in x-direction. The fits for the y-direction give an angle α = 49° in the near field and α = -45° in the far field and the product

$$\Delta^2(y_1 - y_2) \cdot \Delta^2(p_{y,1} - p_{y,2}) = 0.32 \cdot \hbar^2.$$

Using the covariance matrix, the evaluation yields for the y-direction $\sigma_{y,in} = 41.5$ μm $\sigma_{py,in} = 25100$ ℏ/m and correlation coefficients of $\rho_y = 0.45$ and $\rho_{p_y} = -0.86$ leading to a product of $0.32\ \hbar^2$.

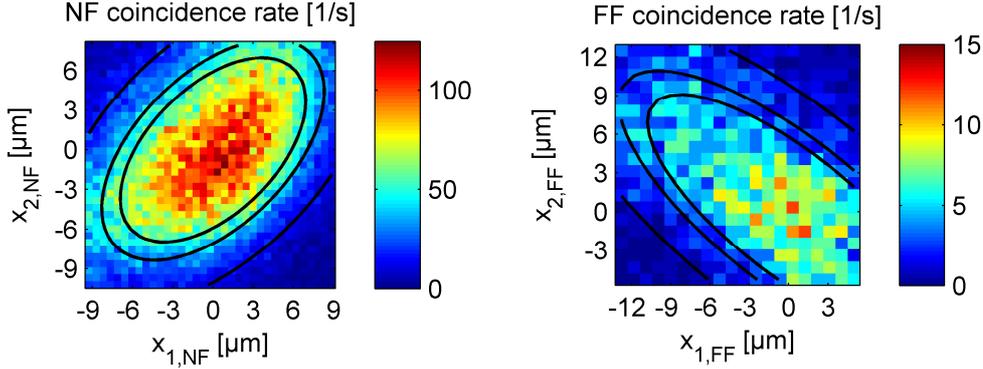

*Fig. 6: Measured coincidence rates of biphoton beam in x-direction detected in a spatially resolved Hanbury-Brown and Twiss setup (see Fig. 4). Measurement in imaged near field plane (left) and in imaged far field plane (right). The contour lines give the 2d-fit-result for the ½, 1/e, and 1/e² -level of the coincidence rate.*

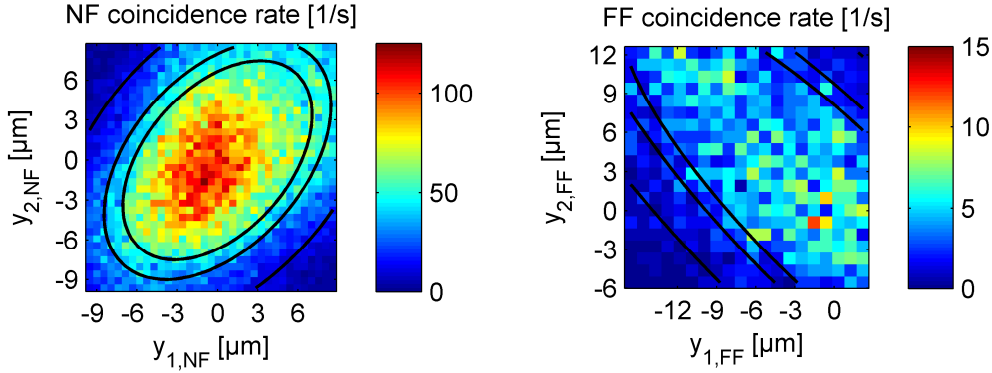

*Fig. 7: Measured coincidence rates of biphoton beam in y-direction detected in a spatially resolved Hanbury-Brown and Twiss setup (see Fig. 4. Measurement in imaged near field plane (left) and in imaged far field plane (right). The contour lines give the 2d-fit-result for the ½, 1/e, and 1/e² -level of the coincidence rate.*

As pointed out in section 2, the observable correlation strength is limited by the number of higher order spatial modes that can be detected with the setup. In addition the near field uncertainty is limited by the thickness of the nonlinear crystal. The longer the crystal the more the origin of signal and idler photon gets blurred if the position of the photon is observed in one specific plane. This is simply because the photons have an anti-correlation in their transverse momentum so that after a certain longitudinal distance their transverse separation increases. This limitation, due to the crystal thickness L with a refractive index of the nonlinear crystal $n_{nlc}$, can be expressed by $\Delta(x_1 - x_2) > \Theta_j/(2n_{nlc}) \cdot L/2$. In our case this amounts to about 1 μm. Whereas the limitation of $\Delta(x_1 - x_2)$ because of the angular spread (see section 2) results to 19 µm in the crystal plane so that the impact due the crystal thickness becomes negligible in this case. The transverse momentum uncertainty $\Delta(p_{x,1} + p_{x,2})$ limited by the pump spot size amounts to 12.500 $\hbar/m$. The product of the latter two contributions creates a limitation for the correlation strength of the variances of 0.056 $\hbar^2$ which is around three times smaller than the product we measured. There are two major reasons why we did not reach this boundary. One is the fiber mode field diameter of 5.3 µm of the fiber probes. Taking this resolution limiting size in account by deconvolution one could extrapolate a measurable product of 0.095 $\hbar^2$ instead of 0.16 $\hbar^2$. The second reason lies in the pointing stability of the setup during the measurement duration. Due to fluctuations of the laser itself, temperature drifts and fluctuations on the table, the remaining difference can be easily explained.

## 5. Conclusion

We report on the characterization of space momentum entanglement in two transverse dimensions for photons generated from type II parametric down conversion in a BBO crystal. The results of our correlation characterization can be used to proof non-separability by testing for the criterion of Mancini et. al. [21]. This criterion is strongly violated showing the non-classical character of the characterized photon correlations. Using the $M^2$-number known as characterization measure of the multimode character of laser beams and covariance matrices a general criterion for optical spatial entanglement is expressed. Applying these criteria, optimal conditions for an optical source for spatial entanglement and its observation are expressed with special emphasis on the spatial multimode detection of such an entangled biphoton state. So far the limits of the degree of our entanglement stem from the limited transversal width of the parametric down conversion source and the limited width of apertures on the propagation to the photon detectors of our characterization setup. The thickness of the nonlinear crystal has not been limiting so far, but in principle poses a limit for high fidelity EPR-correlations from parametric down conversion. Addressing the correlation of specific spatial modes is on target.

## Acknowledgement

Robert Elsner is acknowledged for strong support in programming the fiber stages. Fruitful discussions on the concepts of entanglement characterization with Jens Eisert and Carsten Henkel are gratefully acknowledged.